\documentclass[fleqn,usenatbib]{mnras}

\usepackage[T1]{fontenc}

\DeclareRobustCommand{\VAN}[3]{#2}
\let\VANthebibliography\thebibliography
\def\thebibliography{\DeclareRobustCommand{\VAN}[3]{##3}\VANthebibliography}

\usepackage{graphicx}	
\usepackage{amsmath}	
\usepackage{amssymb}	

\usepackage{amssymb}

\usepackage{newtxmath}

\title[Gamma rays from colliding HVCs]{Gamma rays from reaccelerated cosmic rays in high velocity clouds colliding with the Galactic disk}

\author[M. V. del Valle]{
Maria V. del Valle$^{1}$\thanks{E-mail: mvdelvalle@usp.br (M.V.d.V.)}
\\
$^{1}$Instituto de Astronomia, Geofísica e Ciências Atmosféricas, Rua do Matão, 1226 - Cidade Universitária - 05508-090 São Paulo-SP - Brasil
}

\date{Accepted XXX. Received YYY; in original form ZZZ}

\pubyear{2021}

\begin{document}
\label{firstpage}
\pagerange{\pageref{firstpage}--\pageref{lastpage}}
\maketitle

\begin{abstract}
 
High velocity clouds moving toward the disk will reach the Galactic plane and will inevitably collide with the disk. In these collisions a system of two shocks is produced, one propagating through the disk and the other develops within the cloud. The shocks produced within the clouds in these interactions have velocities of hundreds of kilometers per second. When these shocks are radiative they may be inefficient in accelerating fresh particles, however they can reaccelerate and compress Galactic cosmic rays from the background. In this work we investigate the interactions of Galactic cosmic rays within a shocked high velocity cloud, when the shock is induced by the collision with the disk. This study is focused in the case of radiative shocks. We aim to establish under which conditions these interactions lead to significant nonthermal emission, especially gamma rays. We model the interaction of cosmic ray protons and electrons reaccelerated and further energized by compression in shocks within the clouds, under very general assumptions. We also consider secondary electron-positron pairs produced by the cosmic ray protons when colliding with the material of the cloud. We conclude that nearby clouds reaccelerating Galactic cosmic rays in local shocks can produce high-energy radiation that might be detectable with existing and future gamma-ray detectors. The emission produced by electrons and secondary pairs is important at radio wavelengths, and in some cases it may be relevant at hard X-rays. Concerning higher energies, the leptonic contribution to the spectral energy distribution is significant at soft gamma rays. 
 
\end{abstract}




\section{Introduction}
High-velocity clouds (HVCs) consisting of cold gas are known to be moving in the Galactic halo at anomalous velocities. By convention HVCs have deviation velocities $V_{\rm d} > 90$\,km\,s$^{-1}$, where this velocity is the difference between the observed velocity of the gas and the maximum velocity expected from a simple model of differential Galactic rotation  \citep[see,][]{1991A&A...250..499W}. There are three possible origins for these clouds: the so-called \emph{Galactic fountains} produced by supernovae, tidal streams such as the \emph{Magellanic stream} and low-metalicity gas accreting onto the Milky Way \citep[e.g.,][]{2013pss5.book..587W}. Hence the HVC class shares some similarities between individuals but presents a wide range of properties.
 
A large fraction of HVCs are inferred to be moving toward the Galactic disk so they will reach the Galactic plane at some time and will inevitably collide with the disk \citep[e.g.,][]{2012ARA&A..50..491P}. The impact of these clouds with the gas in the disk  {\rm can dissipate} a large amount of energy, between $10^{47}-10^{52}$\,erg for each cloud \citep{1981A&A....94..338T}. The super massive HVC called the \emph{Smith Cloud}, for example, is thought to have collided with the Galactic disk approximately 70\,Myr ago \citep{2008ApJ...679L..21L}.

The collision with the disk produces a system of two shocks, propagating through the disk and the cloud \citep{1981A&A....94..338T}. Under some conditions, these shocks could be the site of particle acceleration and the accelerated particles might produce nonthermal emission \citep[e.g.,][]{Inoue_2017,2018MNRAS.475.4298D}. Also the clouds can act as passive sources for the cosmic rays, producing gamma rays from neutral pion decay as observed in molecular clouds.
 
The idea that HVCs might produce gamma emission is not new \citep[see][]{1983SSRv...36...93M}. HVCs as potential sites of long-lasting shocks which, like supernova shocks, can serve as sites for particle acceleration in the gaseous halo was proposed by \citet{1977ApJ...215..208H}. \citet{1997A&A...321..288B} studied the possibility that  extended MeV emission detected by COMPTEL may be associated with high-velocity clouds. In their model electrons accelerated in local shocks produce the emission via relativistic Bremsstrahlung \citep[see also][]{2001ESASP.459..123B}. 

Aiming to trace the distribution of cosmic ray nuclei in the Galactic halo for the first time, \citet{2015ApJ...807..161T} used 73 months of data from the {\it Fermi} LAT to search for gamma-ray emission produced by cosmic ray interactions in high- and intermediate-velocity clouds (IVCs). They considered targets located within 7\,kpc above the Galactic plane{, detecting} gamma-ray emission from IVCs for the first time. 

The H.E.S.S. source J1503-582, detected up to $\sim$ 20 TeV \citep{2008AIPC.1085..281R} and its sub-TeV counterpart \citep{2016ApJS..222....5A} were claimed to be associated with a HVC interacting with the Galactic disk \citep{Inoue_2017}. The emission in the TeV regime should be of hadronic nature. These clouds, less dense than molecular clouds, need a power in relativistic protons much in excess of that from the cosmic ray background in order to produce significant gamma rays.

Recently, we modeled the nonthermal emission produced in the collision of high-velocity clouds with the Galactic disk \citep{2017AIPC.1792d0007M,2018MNRAS.475.4298D}. We considered the case of an adiabatic shock accelerating protons and electrons. We found that in a shocked cloud significant synchrotron radio emission is produced along with soft gamma rays. The impact of the clouds with the Galactic disk releases a great amount of energy, we have shown that under some reasonable conditions a fraction of that energy can be converted into nonthermal radiation and energetic particles; \citet{Inoue_2017} arrived at similar conclusions.

The shocks formed within the clouds after the collisions with the Galactic disk have velocities of hundreds of kilometers per second  and will be radiative in many cases \citep[e.g.,][]{2018MNRAS.475.4298D}. Cosmic rays  from the so called \emph{Cosmic-ray Sea} permeate  the cloud, and they can be reaccelerated at the shocks independent of their energy. Slow shocks can contribute  to nonthermal particle reacceleration, while they are not expected to be efficient at accelerating thermal particles \citep{1982ApJ...260..625B,2016A&A...595A..58C}. Also the reaccelerated particles and \emph{fresh} accelerated particles could experience further energization as a result of shock compression \citep[e.g.,][]{2011A&A...527A..99E}. In radiative shocks the gas is compressed to values much greater than the factor of 4 produced by adiabatic shocks, hence re-energizing the relativistic particles significantly. This reacceleration+compression scenario has been studied in molecular clouds in the presence of slow shocks \citep[e.g.,][]{Uchiyama_2010,2015ApJ...806...71L,2016A&A...595A..58C}.

In this work we explore the physical situation described above in which cosmic rays from the Galactic background interact within HVCs colliding with the Galactic disk. Particles could suffer acceleration and compression in a radiative shock induced by the cloud collision. The high-energy particles are expected to cool down, emitting nonthermal photons. In this work we estimate the produced nonthermal emission for a wide range of parameters, and possible physical situations/conditions. We also analyze in which cases the gamma-ray emission might be detectable by current and future detectors.

In the next Section we present the model we implement to estimate the population of cosmic rays in the cloud considering acceleration and reacceleration in the induced shock. Then in Sect.\,\ref{results} we show the results of the nonthermal emission produced by the different particle populations, and we also present an extensive discussion. Finally, in Sect.\,\ref{conclusion} we offer our conclusions and remarks.

\section{Acceleration and emission model}\label{modelling} 

The cloud collision with the disk produces a shock with a velocity $\sim$ $V_{\rm c}$, being this last the cloud's velocity.
These velocities are of the order of 100\,km\,s$^{-1}$. Here we assume a velocity range between 100 and 300\,km\,s$^{-1}$. At these velocities shocks tend to be radiative. In these shocks the radiative cooling of the shock-heated gas is efficient enough to be dynamically important. Hence in a radiative shock the cooling time is shorter than the dynamical time. 

The condition for a shock to be radiative can be translated in terms of the cooling column density $N_{\rm cool}$, defined as $N_{\rm cool} = n_{\rm c}\,V_{\rm s}\,t_{\rm cool}$, where $t_{\rm cool}$ is the cooling time $t_{\rm cool} = {3n_{\rm c}k_{\rm B}T}/
({n_{\rm c}^{2}\Lambda(T)})$, with $k_{\rm B}$ the Boltzmann's constant and $\Lambda(T)$ the cooling function. Here $T$ is the temperature immediately after the shock front, { which is} $T = 1.38\times 10^{5}\left({V_{\rm s}}/{100\,{\rm km}\,{\rm s}^{-1}}\right)^2\,$K for a fully ionized gas. We use the cooling function for optically-thin plasma in collisional ionization equilibrium and solar metalicity \citep[e.g.,][]{1976ApJ...204..290R} that can be parametrized as a power-law $\Lambda(T) \sim 7\times10^{-19}T^{-0.6}$\,erg\,cm$^3$\,s$^{-1}$, in the temperature range of interest. Then we can impose a lower limit in the cloud's density:
\begin{equation}\label{eq4}
n_{\rm c} \geq 4\times 10^{-3}\left( \frac{V_{\rm s}}{100\,{\rm km}\,{\rm s}^{-1}} \right)^{4}\left( \frac{R_{\rm c}}{20\,{\rm pc}} \right) \,{\rm cm}^{-3},
\end{equation}

\noindent which is not a very restrictive limit, even for an extreme case of $V_{\rm s} \sim 300$\,km\,s$^{-1}$ which gives $n_{\rm c} \geq 0.3$\,cm$^{-3}$. As mentioned previously HVCs have multiple origins, hence differences in metalicity from source to source can be expected; these differences impact the cooling function and could change the above condition.

In an adiabatic strong shock the compression factor $S$, defined as the ratio of the postshock density to the upstream density, reaches a value $\sim$ 4. For radiative shocks however the value of $S$ can be much higher. The compression is halted by the magnetic pressure in a magnetized plasma. In order to obtained the compression factor $S$ we equate the shock ram pressure $n_{c}\,\mu_{\rm H}V_{\rm s}^2$ to the magnetic pressure $B_{\rm m}^2/8\pi$ \citep{1982ApJ...260..625B}. Assuming that the magnetic field is randomly directed ahead of the shock, the compressed magnetic field becomes $B_{\rm m} = \sqrt{2/3}\,SB_{\rm c}$, being $B_{\rm c}$ the magnetic field of the cloud. The compression factor yields:
\begin{equation}\label{S}
S = 18.8 \left(\frac{n_{c}}{\rm cm^{-3}} \right)^{1/2} \left( \frac{V_{\rm s}}{\rm 100\,km\,s^{-1}}\right)\left(\frac{B_{\rm c}}{5\,\mu{\rm G}} \right)^{-1}.  
\end{equation} 

Magnetic field measurements from HVCs are scarce. From 21-cm Zeeman measurements, for example, values of  $\sim$ 10\,${\mu}$G  in a core of complex A were obtained \citep{1991A&A...245L..17K}. Using Faraday rotation measurements \citet{2013ApJ...777...55H} estimated a lower limit for the line-of-sight component of $\sim$ 8\,$\mu$G for the Smith Cloud. Most recently \citet{2019ApJ...871..215B} found an upper limit of $\sim$ 5\,$\mu$G for the same source. Here we adopt a value of $B_{\rm c} = 5$\,$\mu$G. For this magnetic field we obtain $S \sim 20$ for $V_{\rm c} = 100$\,km\,s$^{-1}$ and $S \sim 60$ for $V_{\rm c} = 300$\,km\,s$^{-1}$. 

For estimating the cosmic ray distribution we use the cosmic ray flux measured by Voyager 1 in 2013. In the case of protons it is parametrized as \citep[e.g.,][]{2016Ap&SS.361...48B,2016A&A...595A..58C}:
\begin{eqnarray}\label{cr}
J_{\rm p} = 232.4 \left(\frac{E_{\rm GeV}^{1.02}}{\beta_{p}^2}\right) \times \left( \frac{E_{\rm GeV}^{1.19}+0.77^{1.19}}{1+0.77^{1.19}}\right)^{-3.15}\\ \nonumber
\#{\rm part.}\,{\rm cm^2}\,{\rm s}^{-1}\,{\rm sr}^{-1}\,{\rm erg}^{-1}.
\end{eqnarray}
\noindent $E_{\rm GeV}$ is the proton's kinetic energy in units of Giga-electron-volts, and $\beta_{p}$ is the protons normalized velocity, which depends on the particle's  energy. 

Similarly, for cosmic ray electrons we use the following parametrization \citep{Potgieter_2015}: 
\begin{eqnarray}\label{cre}
J_{\rm e} = 0.21 \left(\frac{E_{\rm GeV}^{-1.35}}{\beta_{e}^2}\right) \times \left( \frac{E_{\rm GeV}^{1.65}+0.6920}{1+0.6920}\right)^{-1.1515} +\\ \nonumber
 1.73\exp\left[4.19-5.40\log(E_{\rm GeV})-8.9E_{\rm GeV}^{-0.64}\right]\\ \nonumber
\#{\rm part.}\,{\rm m^2}\,{\rm s}^{-1}\,{\rm sr}^{-1}\,{\rm MeV}^{-1}.
\end{eqnarray}
\noindent Here $E_{\rm GeV}$ is the electron's kinetic energy in units of Giga-electron-volts, and $\beta_{e}$ is the  normalized velocity.

We also consider secondary pairs produced by proton-proton inelastic collisions with the cloud material. The distribution of pairs depends on the proton distribution and medium density, we follow the approach given in  \citet{2006PhRvD..74c4018K}. 

\subsection{Reacceleration}

The particles from the cosmic ray background can be accelerated at the induced shock via first order Fermi process; a radiative shock with $V_{\rm s}\geq 100$\,km\,s$^{-1}$ can ionize the post and the pre-shock regions \citep[e.g.,][]{1996ApJS..102..161D}, hence we consider no damping of the Alfvén waves caused by neutrals, that can halt shock particle acceleration (see further discussion in Sect.\,\ref{sec:wd}). The resulting cosmic ray distribution as a function of the particles momentum $p$ after the reaccelerarion is given by \citep[e.g.,][]{2004APh....21...45B,2016A&A...595A..58C}:
\begin{equation}
f^{\rm CR}_{\rm reacc}(p) = \alpha \left(\frac{p}{p_{\rm m}}\right)^{-\alpha}\int_{p_{\rm m}}^{p}\frac{{\rm d}p'}{p'}\left(\frac{p'}{p'_{\rm m}}\right)^{\alpha}f^{\rm CR}(p');
\end{equation}
\noindent the value of $\alpha$ is related to the shock compression before cooling (adiabatic), we consider a value of $\alpha = 4$; $p_{\rm m}$ is a minimum momentum value, we take $p_{\rm m} = 50\,{\rm MeV}/c$. Finally, $f^{\rm CR}(p)$  is related to  $J(E)$ through: 
\begin{equation}
4 \pi p^2 f^{\rm CR}(p) {\rm d}p = \frac{4\pi c}{\beta}J(E){\rm d}E. 
\end{equation} 

The reacceleration of the background cosmic rays increases the momentum per particle until a maximum momentum/energy, hence we multiply $f^{\rm CR}_{\rm reacc}(p)$ by a factor $\exp \left( {- p / p_{\rm max}}  \right)$. The maximum energy is estimated in Sect.\,\ref{emax}.

\subsection{Compression}\label{compress}

The cosmic rays can experience further energization  due to adiabatic compression, as the gas density increases until the pressure is magnetically supported (see Eq.\,(\ref{S})). The compression shifts the distribution to higher energies, because each
particle gains energy as $p \rightarrow \left(S/r_{\rm sh}\right)^{1/3}\,p$, with $r_{\rm sh}$ the shock adiabatic compression here considered to be equal to 4. Also the particle density increases by a factor $S/r_{\rm sh}$ \citep[see][]{1982ApJ...260..625B,Uchiyama_2010}. Therefore after the compression:
\begin{equation}
f^{\rm CR}_{\rm compress}(p) = \left(\frac{S}{r_{\rm sh}}\right)^{2/3} \times f^{\rm CR}_{\rm reacc}\left([S/r_{\rm sh}]^{-1/3}\,p\right).
\end{equation} 

The reacceleration and compression of cosmic rays in radiative shocks is treated here as a two-step process, however the physics of particle acceleration in radiative shocks is uncertain and alternative scenarios have been discussed. For example, \citet{2006MNRAS.371.1975Y} argued that particle acceleration in the context of SNR radiative phase may lead to a hard cosmic ray spectrum with index $< 2$. 

\subsection{Acceleration of fresh particles}\label{acc}

Fresh particles can also be accelerated at the shock by diffusive shock acceleration. We consider both electrons and protons, assuming a proton-to-electron power ratio  $a \sim 100$\footnote{It should be noticed that this value of the proton-to-electron power ratio is the one observed in cosmic rays at Earth and it might not hold at the acceleration site}. From this mechanism a power-law distribution is expected with index $\alpha_{\rm f} \sim 4$, i.e.:
\begin{equation}
f_{\rm f} = {K}_{\rm f} \left(\frac{p}{p_{\rm inj}}\right)^{-\alpha_{\rm f}},
\end{equation}  
\noindent with  $p_{\rm inj}$ the injection momentum corresponding to the energy $E_{\rm inj} \approx 4.5\,E_{\rm sh}$, and $E_{\rm sh} = \frac{1}{2}m_{p}V_{\rm s}^{2}$ \citep{2014ApJ...783...91C}.  ${K}_{\rm f}$ is the normalization factor such that
\begin{equation}
K_{\rm f} = \frac{\eta_{\rm CR} \rho_{\rm c} V_{\rm s}^{2}}{4/3 \pi c \int \left(p/p_{\rm inj} \right)^{-\alpha_{\rm f}}p^3 \beta_p {\rm d}p},
\end{equation} 
\noindent here $\eta_{\rm CR}$ is the efficiency of particle acceleration. In non-relativistic shocks $\eta_{\rm CR} \sim 0.1$ \citep[e.g.,][]{2014ApJ...783...91C} and can be as high as 0.25 in Supernova remnants. For the radiative shocks in this work we adopt a more conservative value of $\eta_{\rm CR} = 0.025$ \citep[see,][]{2018MNRAS.479..687S}.

\subsection{Maximum energy}\label{emax}
Particles will be accelerated (or reaccelerated) in the shock up to a maximum energy $E_{\rm max}$. In these sources the energy losses for high-energy protons are not dominating and the maximum achievable energy is given by the lifetime of the accelerator, this is $t_{\rm acc} < t_{\rm int}$, with $t_{\rm int}$ the interaction time  $t_{\rm int} = \zeta R_{\rm c}/V_{\rm s}$, with $\zeta < 1$ which is a fraction of the shock crossing time. Also given that the shock itself ionizes the preshock region, wave damping is not expected to halt acceleration in this scenario. 
The acceleration time is given by: 
\begin{equation}
t_{\rm acc} = \frac{D(E)}{V_{\rm s}^2},
\end{equation}
\noindent with $D(E)$ the particles spatial diffusion coefficient. An order of magnitude estimation for the diffusion coefficient as a function of the cloud's turbulence is given by \citep[e.g.,][]{2016A&A...595A..58C}: 
\begin{equation}\label{emax1}
D(E) \approx \frac{1}{3}r_{\rm L}c \frac{B_{0}^2}{\delta B^2}\left(\frac{L_{\rm c}}{r_{\rm L}}\right)^{k-1},
\end{equation}
\noindent with $r_{\rm L} = \frac{E / {\rm eV}}{300\,B / {\rm G}}$\,cm the particle's Larmor radius, and $L_{\rm c}$ the coherence length of the magnetic field. Assuming  a Kolmogorov spectrum for the turbulence, i.e. $k = 5/3$, as observed in the interstellar medium \citep{1995ApJ...443..209A,2010ApJ...710..853C}. Also we assume  ${B_{0}^2}/{\delta B^2} \approx 1$ for $k_{0} = 1/L_{\rm c}$, with $L_{\rm c}$  the injection scale of turbulence,  then the maximum energy is given by:
\begin{eqnarray}\label{emax2}
E_{\rm max} \approx && 4\, {\rm GeV} \left(\frac{\zeta}{0.1}\right)^3 \left(\frac{B_{0}}{5\,\mu{\rm G}}\right)\left(\frac{V_{\rm s}}{100\,{\rm km}\,{\rm s}^{-1}}\right)^3  \times \nonumber \\ 
&& \left(\frac{R_{\rm c}}{20\,{\rm pc}}\right)^3 \left(\frac{L_{\rm c}}{0.1\,{\rm pc}}\right)^{-2}.
\end{eqnarray}
\noindent Therefore  $E_{\rm max} \approx 4\, {\rm GeV}$ for $V_{\rm s} = 100\,$km\,s$^{-1}$ and  $E_{\rm max} \approx 36\, {\rm GeV}$ for $V_{\rm s} = 300\,$km\,s$^{-1}$. 

Maps of the HVCs have shown small-scale structure down to the scale of the angular resolution. Different approaches have been used to numerically characterize these structures. Also, some models are inspired by theoretical models of the ISM (interstellar medium), especially the idea that the structures are generated by turbulence \citep{2013pss5.book..587W}. Therefore these clouds seems to have some level of turbulence, but  estimations of  $L_{\rm c}$ for HVCs are far to be obtained. Here we assume a value $L_{\rm c} \sim 0.1$\,pc, which is the turbulence correlation length in molecular clouds \citep{2009ApJ...706.1504H}. 

On the other hand, near the shock particle acceleration is expected to be efficient if the spatial diffusion of the particles occurs in the Bohm regime, with $D(E) \approx D_{\rm B}(E) = 1/3\,r_{\rm L} c$. In this case the maximum achievable energy in the shock acceleration process is much higher, given by: 
\begin{eqnarray}
E_{\rm max,\, B} \approx 9\,{\rm TeV} \left(\frac{\zeta}{0.1}\right) \left(\frac{V_{\rm s}}{100\,{\rm km}\,{\rm s}^{-1}}\right)  \left(\frac{R_{\rm c}}{20\,{\rm pc}}\right)\left(\frac{B_{0}}{5\,\mu{\rm G}}\right),
\end{eqnarray} 
\noindent then $E_{\rm max,\,B} \approx 9\, {\rm TeV}$ for $V_{\rm s} = 100\,$km\,s$^{-1}$ and $E_{\rm max,\,B} \approx 27\, {\rm TeV}$ for $V_{\rm s} = 300\,$km\,s$^{-1}$. In the absence of a better estimation of the diffusion coefficient we consider both cases $E_{\rm max}$ and $E_{\rm max,\,B}$.
   
\subsection{Particle distribution}

In order to obtain the final particle distribution we solve the following transport equation: 
\begin{equation}
\frac{\partial N(E,t)}{\partial t} = \frac{\partial}{\partial E} \left[ b(E) N(E,t)\right] + Q(E),
\end{equation}
\noindent $b(E) \equiv {\rm d}E/{\rm d}t$ are the energy losses. 

$Q$ is the injection function given by \citep[e.g.,][]{2016A&A...595A..58C} in the case of primary particles:
\begin{equation}\label{trans}
Q(E) = S^{-1}\frac{1}{t_{\rm int}}N'(E), 
\end{equation}
and $N'(E) = 4 \pi p^3 f(p)\,{\rm d}p/{\rm d}E$. 

The stationary solution of the above equation is given by:
\begin{equation}
N(E) = \bigg|\frac{{\rm d}E}{{\rm d}t}\bigg|^{-1}\int_{E}^{\infty} Q(E')\,{\rm d}E.
\end{equation}
\noindent  
The relevant energy losses for protons are given by ionization losses, important at the lower energies and proton-proton inelastic collisions, which are relevant at higher energies. 

In the case of electrons the relevant losses are: synchrotron produced by interactions with the local magnetic field, relativistic Bremsstrahlung and inverse Compton (IC) scattering. Being the first two dominating. As targets for the IC we consider the CMB (cosmic microwave background) and FIR (far-infrared interstellar radiation field), with energy densities $U \sim 0.261$\,eV\,cm$^{-3}$ and $U \sim 1$\,eV\,cm$^{-3}$, respectively. We also consider the photons produced by the radiative shock itself following the treatment presented in \citet{10.1093/mnras/stu844}. We obtain  $U \sim 0.0057$\,eV\,cm$^{-3}$ at $T \sim 5\times10^{4}$\,K for a $V_{\rm s} = 100$\,km\,s$^{-1}$ and $U \sim 0.157$\,eV\,cm$^{-3}$ at $T \sim 6.1\times10^{5}$\,K for the fastest shock.

\section{Results and discussion}\label{results}

   \begin{figure*}
   \centering
   \includegraphics[width=0.3\linewidth, trim=1cm 0.8cm 1.5cm 0.8cm, angle=270]{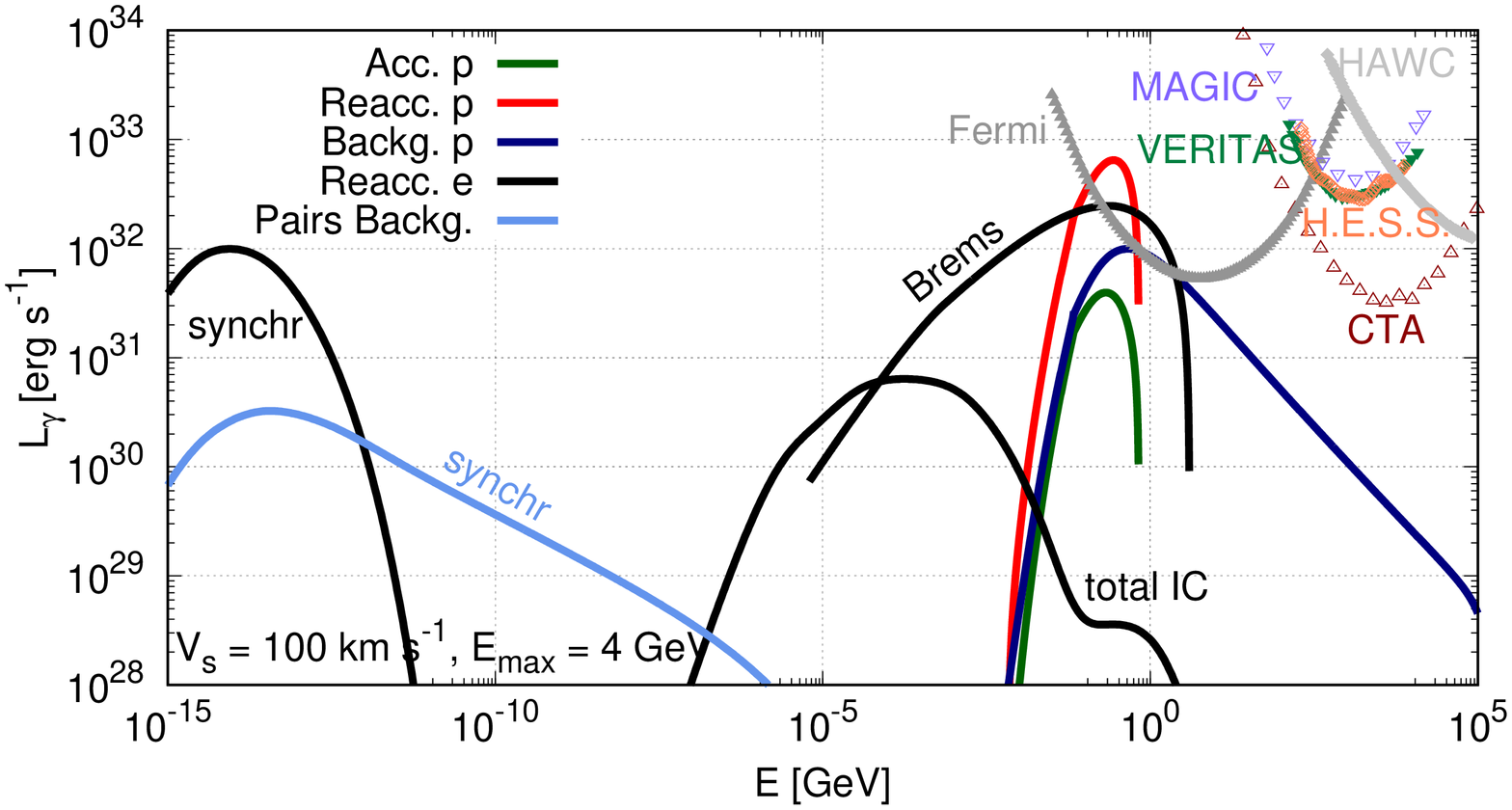}
     \includegraphics[width=0.3\linewidth, trim=1cm 0.8cm 1.5cm 0.8cm, angle=270]{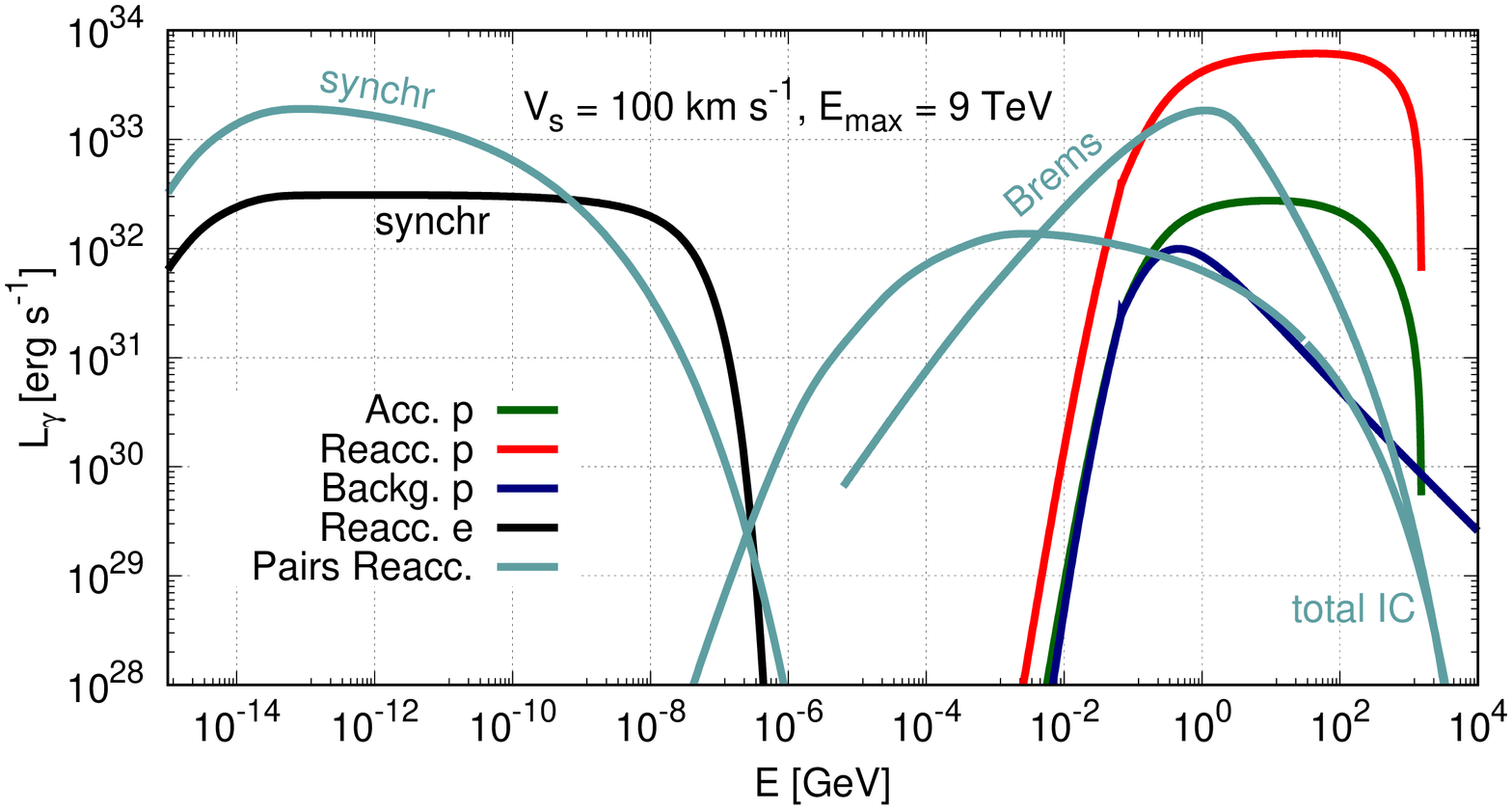}
      \caption{Nonthermal SED expected from the reacceleration of background cosmic rays, secondary pairs and fresh acceleration of particles via first order Fermi acceleration in a shock with $V_{\rm s} =100$ in a HVC. Two cases for $E_{\rm max}$ are considered: 4\,GeV (left panel) and 9\,TeV (right panel). { In the left plot are also shown the sensitivity curves for CTA-SOUTH, H.E.S.S., VERITAS, MAGIC, HAWC and {\it Fermi} LAT. See further details in the text.}}
         \label{v100}
   \end{figure*}

   \begin{figure*}
   \centering
   \includegraphics[width=0.3\linewidth,  trim=1cm 0.8cm 1.5cm 0.8cm, angle=270]{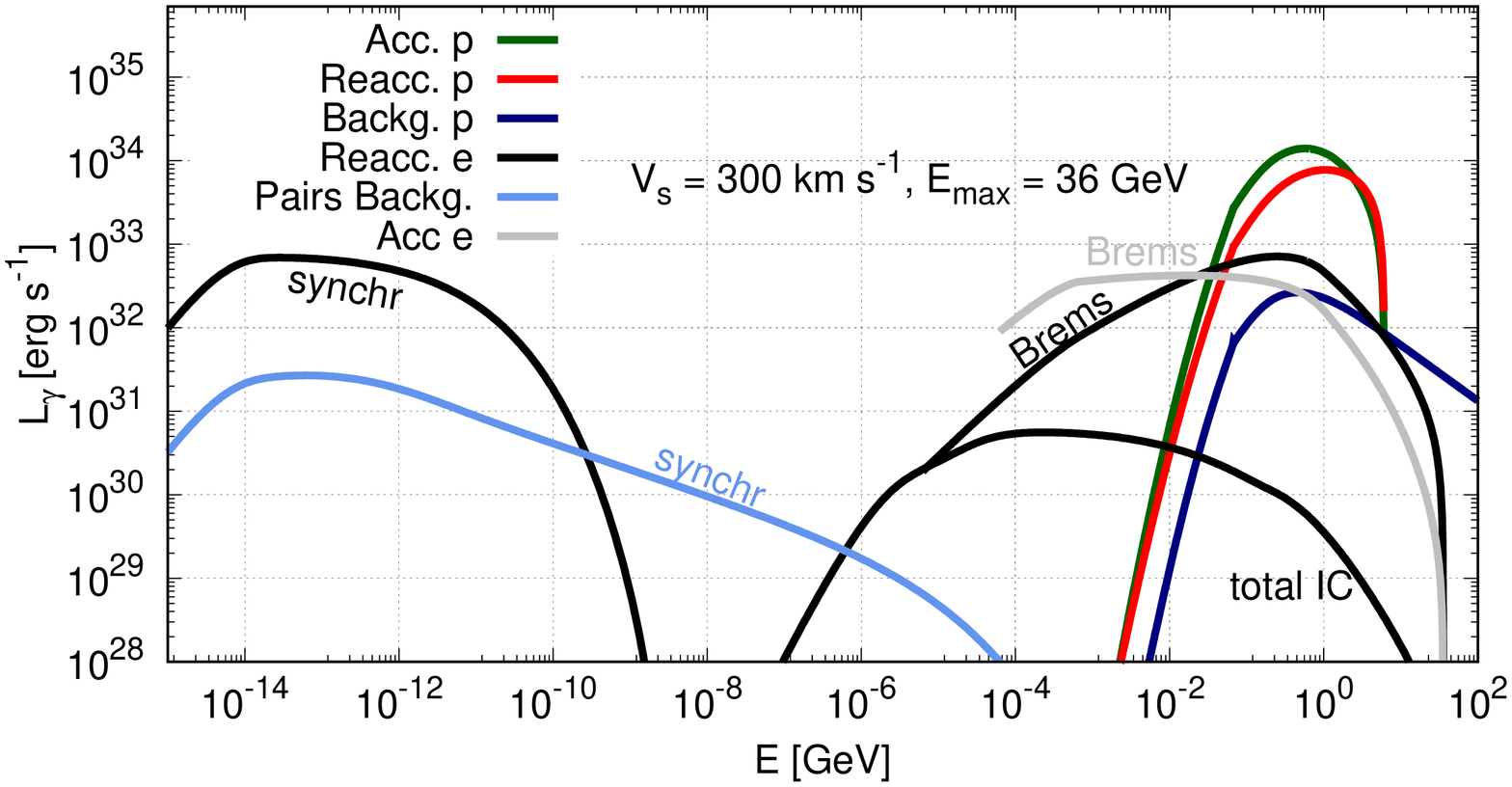}
      \includegraphics[width=0.3\linewidth,  trim=1cm 0.8cm 1.5cm 0.8cm, angle=270]{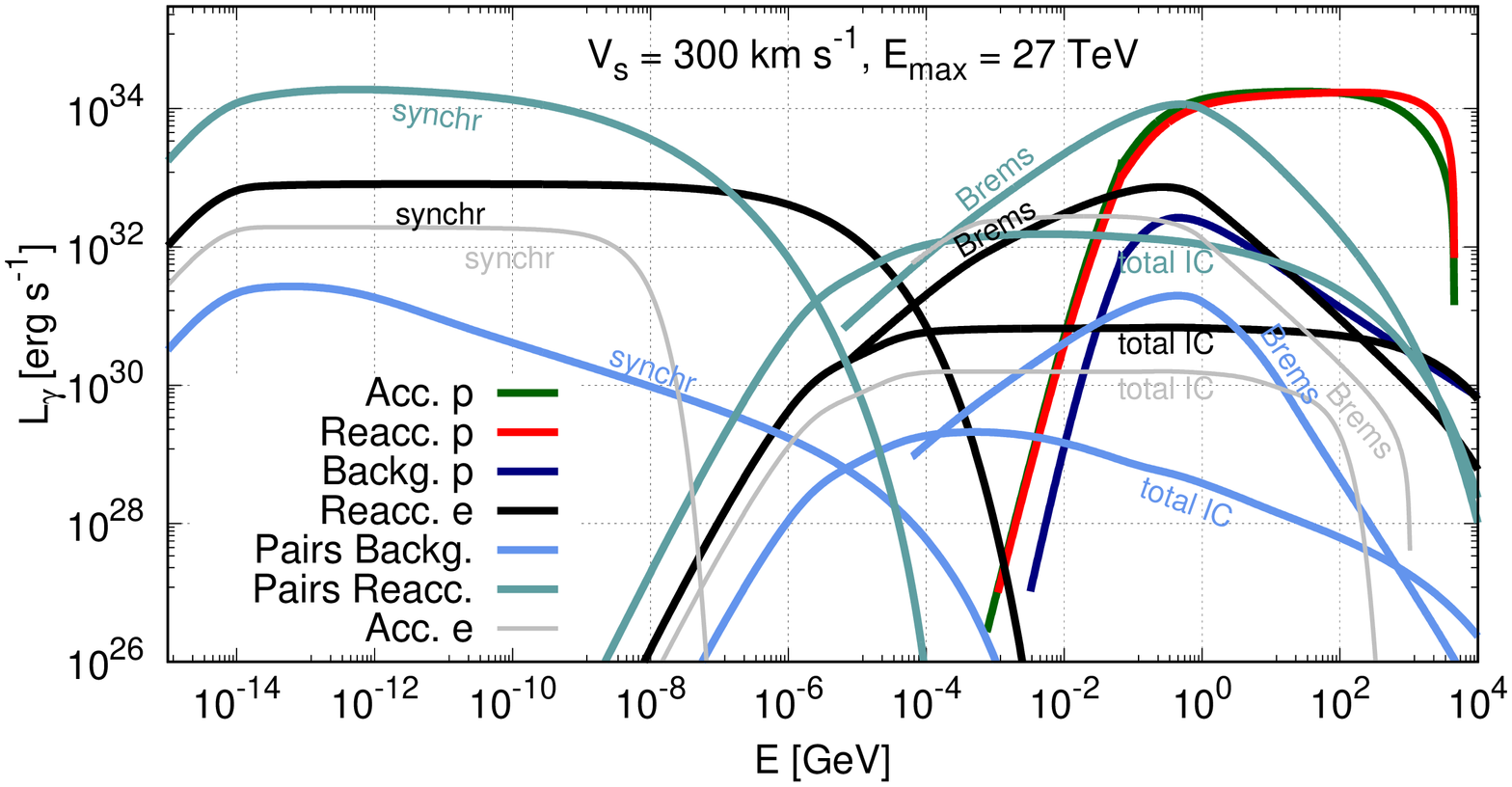}
      \caption{{ Nonthermal SED} expected from the reacceleration of background cosmic rays,  secondary pairs and fresh acceleration of particles via first order Fermi acceleration in a shock with $V_{\rm s} = 300$ in a HVC. Two cases for $E_{\rm max}$ are considered: 36\,GeV (left panel) and 27\,TeV (right panel).}
         \label{v300}
   \end{figure*}

The protons interact with the material of the cloud producing gamma rays through neutral pion decay  (further byproducts are also produced, see Sect.\,\ref{neutrinos}). We calculate this emission in the delta-function  approximation \citep[e.g.,][]{1991Ap&SS.180..305A}. The leptons will produce nonthermal emission through synchrotron, IC and relativistic Bremsstrahlung. For computing the synchrotron and IC components we use the \emph{Naima} Python package \citep{naima}; in computing the relativistic Bremsstrahlung we consider the fully ionized case. 

We calculate the emission from three high-energy proton populations: i) reaccelerated cosmic ray protons, ii) freshly accelerated protons, and iii) cosmic ray protons from the background.  We also compute the emission produced by four leptonic populations: i) the background cosmic ray electrons reaccelerated at the shock, ii) freshly accelerated electrons; plus two secondary-pairs populations produced by:  iii) the background cosmic ray protons and  iv) the reaccelerated cosmic ray protons. All populations undergo re-energization due to compression. We estimate the maximum energies achievable in the diffusive shock acceleration by two methods (see Sect.\,\ref{emax}), obtaining:   4\,GeV and 9\,TeV. The actual maximum energy might lie between these two values. We show only the dominating components in the SEDs (spectral energy distribution) but in one case in which all contributions are shown for completeness.

The Figure\,\ref{v100}  shows the nonthermal emission expected for $V_{\rm s} = 100$\,km\,s$^{-1}$. Also shown are the sensitivity curves of the Cherenkov detectors: MAGIC, VERITAS, H.E.S.S. and the future CTA-SOUTH all for 50\,h of integration, and HAWC for 5\,yr,  together with the sensitivity curve of the ${\it Fermi}$ LAT for 10\,yr \citep[see e.g,][]{2019scta.book.....C}. These values correspond to a source located at $d = 2$\,kpc, assuming a point source. Note that these sensitivity curves are indicative only, in general these values depend on many factors, such as the specific position of the source, the background level, and source extension; also the curves correspond to long periods of integration that might be prohibitive in many cases. Hence, the detectability prediction needs a careful analysis for each specific source, which is not the purpose of the current work.

In the case of a lower maximum energy (left panel) the luminosity reaches values of the order of 6$\times 10^{32}$\,erg\,s$^{-1}$ at $\sim$ 0.1\,GeV. The contribution from the freshly accelerated particles is negligible in the presence of reaccelerated cosmic rays. The compressed cosmic rays from the background dominate the spectrum for $E_{\gamma} > 0.6$\,GeV. For a cloud at a distance of $2$\,kpc $F_{\gamma} \sim 2.6\times 10^{-10}$\,cm$^{-2}$\,s$^{-1}$ at $E_{\gamma} = 0.3$\,GeV, this is detectable by {\it Fermi} (see Figure\,\ref{v100}). The accelerated background protons having $E_{\rm max} \sim$\,4\,GeV do not produce a significant population of secondaries and therefore the emission they produce is negligible (the same happens in the case of a higher shock velocity and $E_{\rm max} \sim$\,36\,GeV). The contribution from the cosmic ray electrons dominates the spectrum at soft gamma-rays and at a short range of energies around $0.8$\,GeV. After a couple of GeVs the emission from the background cosmic ray protons dominates the SED. Radio emission is also produced by synchrotron radiation, reaching luminosities of almost $10^{32}$\,erg\,s$^{-1}$; the synchrotron component from the cosmic ray background electrons is broad in energy, however the luminosity is bellow $10^{30}$\,erg\,s$^{-1}$.
 
For a higher maximum energy (right panel) the luminosity reaches much higher values $\sim$  6$\times 10^{33}$\,erg\,s$^{-1}$, this is because much of the high-energy protons producing the neutral pions for gamma-ray emission are not reaccelerated in the case of  a low $E_{\rm max}$. The contribution from the compressed cosmic ray background protons dominates the SED only after $E_{\gamma} > 1.3$\,TeV. At 100 GeV $F_{\gamma} \sim 6.5\times 10^{-11}$\,cm$^{-2}$\,s$^{-1}$, which is above CTA's sensitivity and marginally over that of MAGIC. The emission from secondary pairs from the reaccelerated protons dominates at soft gamma rays, with $\sim$ $10^{32}$\,erg\,s$^{-1}$ around MeV energies. The radio emission reaches values above $10^{33}$\,erg\,s$^{-1}$, also dominated by these pairs; a small region of the SED around 10\,eV is dominated by the emission produced by cosmic ray electrons. 
   
The results for a faster shock, $V_{\rm s} = 300$\,km\,s$^{-1}$, are shown in Figure\,\ref{v300}. Again we consider 2 different maximum energies: 36\,GeV (left panel) and 27\,TeV (right panel, we show all the components in this plot). For this shock the contribution from the compressed cosmic ray protons is negligible. The maximum luminosities at gamma rays do not differ much between the two cases: 1.3$\times 10^{34}$\,erg\,s$^{-1}$ for $E_{\rm max} = 36$\,GeV, and $1.8\times 10^{34}$\,erg\,s$^{-1}$ for the larger maximum energy of 27\,TeV. Naturally this last scenario covers a much wider energy range in the SED and it is most favorable for a detection. An important difference with the slow shock case is the contribution from freshly accelerated protons, with higher shock power to accelerate particles. This component dominates the SED up to 2.5\,GeV in the case of the lower maximum energy, and up to 85\,GeV in the other case. At $E_{\gamma} = 1$\,TeV $F_{\gamma} \sim 3\times 10^{-11}$\,cm$^{-2}$\,s$^{-1}$, a promising value for detection by CTA and existing imaging atmospheric Cherenkov telescopes (IACTs). At these energies the angular resolution of the IACTs, between 0.1 to 0.05$^{\circ}$, might be smaller than the  dimension of the source $\zeta\times$20\,pc, and the hypothesis of point source not longer holds. For a source located at 2\,kpc the corresponding angular size is $L_{\alpha} \sim 0.06^{\circ}$, still a point source for the existing IACTs, but not for CTA, expected to have an angular resolution $dl \sim 0.05^{\circ}$ at 1\,TeV \citep[see,][]{2019scta.book.....C}. In this case the sensitivity is degraded from the point source one by a factor of $L_{\alpha}/dl$, for CTA this factor is 1.2. However this effect can be more dramatic in other cases discussed below. 

Another noticeable difference between the two cases of maximum energy is that the pairs from the accelerated protons dominate the synchrotron output and  they produce soft gamma rays with power above  $10^{32}$\,erg\,s$^{-1}$ in the case of highest energies, while in the other case fresh accelerated electrons dominate the gamma leptonic contribution. The radiation produced by freshly accelerated electrons would dominate the SED in a case with $a \sim 1$. 

The case of a fast shock with $E_{\rm max} \sim 27\,$TeV gives the most significant emission at X-rays, i.e. $\sim$ 1\,keV, giving luminosities of $4\times10^{32}$\,erg\,s$^{-1}$. For a source at $d = 2\,$kpc the flux is $\sim$ $8.4\times10^{-13}$\,erg\,s$^{-1}$\,cm$^{-2}$. {\it XMM-Newton} EPIC-pn point source sensitivity is $\sim$ $5\times10^{-14}$\,erg\,s$^{-1}$\,cm$^{-2}$ for $10^{4}$\,s exposure time, based on a 5$\sigma $ source detection criterion in the energy range $2-10$\,keV, and has an angular resolution of 6 arcsec \citep{2001A&A...365L..51W}. Therefore for an extended source of $L_{\alpha} = 0.06^{\circ}$, the flux sensitivity yields $\sim$ $1.8\times10^{-12}$\,erg\,s$^{-1}$\,cm$^{-2}$, slightly above the emitted flux.

The luminosities at soft gamma rays range from $10^{32}$\,erg\,s$^{-1}$ to $3\times10^{33}$\,erg\,s$^{-1}$ at 10\,MeV. Compton space missions such as AMEGO \citep[All-sky Medium Energy Gamma-ray Observatory,][]{2019BAAS...51g.245M}, e-ASTROGAM \citep{2018JHEAp..19....1D} or GRAMS \citep[Gamma-Ray and AntiMatter Survey,][]{2020APh...114..107A}, which are expected to operate between 200\,keV to 10\,GeV, might be able to detect this emission. However, even for the best scenario ($V_{\rm s} = 300$\,km\,s$^{-1}$ and $E_{\rm max} = 27$\,TeV) a cloud at a $d < 2$\,kpc would be marginally detectable by AMEGO, having a sensitivity $\sim$ $7.7\times10^{-12}$\,erg\,s$^{-1}$\,cm$^{-2}$ at 10\,MeV. 
   
\subsection{Total emission}

The total mass of the modeled cloud in this work is $\sim$ $8\times10^{2}$\,M$_{\odot}$, two orders of magnitude below the expect total mass of a HVC. A spherical cloud of density $n_{\rm c} \sim 1$\,cm$^{-1}$ and total mass $\sim$ $10^{5}$\,M$_{\odot}$ has a radius $\sim$ 88\,pc. In such a large structure alterations are expected, such as the shock slowing down, etc., this is why we consider a smaller cloud for our analysis. In reality the particles accelerated in a smaller volume (here of interaction spatial dimension $\zeta\times$20\,pc) would diffuse into the entire volume of the cloud. This scenario requires different calculations including the propagation of cosmic rays in the cloud in a 2D configuration \citep[e.g.,][]{2015MNRAS.448..207D,2018MNRAS.475.4298D} and will be presented in a future work. 

   \begin{figure}
   \centering
   \includegraphics[width=0.74\linewidth, angle=270]{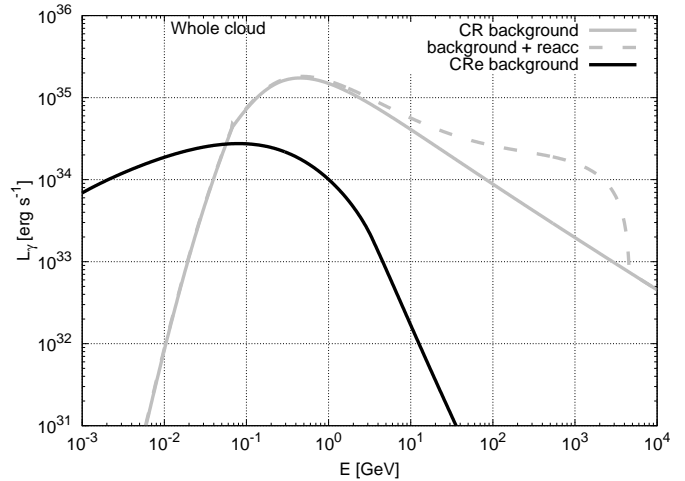}
      \caption{Gamma-ray emission expected from the interaction of cosmic ray protons from the background  in the whole cloud (solid gray line). The dashed gray line shows the sum of the background component plus the contribution from cosmic rays reaccelerated at a 300\,km\,s$^{-1}$ shock.The black line corresponds to the emission produced by cosmic ray electrons from the background.}
         \label{background}
   \end{figure}

We can estimate the emission for the whole cloud ($R_{\rm c} = 88$\,pc, $M_{\rm c} = 10^{5}$\,M$_{\odot}$) interacting with the background of cosmic ray protons (without any reenergization, i.e. a passive source). The result is plotted in Figure\,\ref{background}, gray line. We also show in dashed gray line the sum of the background contribution plus the contribution from reaccelerated cosmic ray protons in a radiative 300\,km\,s$^{-1}$ shock (see Figure\,\ref{v300}, right panel), produced in a fraction of the volume of the cloud. This last contribution significantly enhances the emission above 10\,GeV. Also plotted is the emission from cosmic ray electrons. It is clear that the leptonic contribution to the high-energy SED is adding a component at soft gamma rays.

The gamma-ray flux between 100\,GeV and 1\,TeV is of the order of $2\times10^{-11}$\,\,erg\,s$^{-1}$\,cm$^{-2}$. With an extension at 2\,kpc of $\sim$ 2.5$^{\circ}$ the cloud is an extended source for the gamma-ray detectors (see discussion in Sect.\,\ref{results}). The point source sensitivities of the IACTs would be degraded from the point source one by a maximum factor of 50, corresponding to CTA. Even considering this degradation, a cloud with these characteristics lies above the sensitivity of CTA and of existing IACTs for energies around 100\,GeV. For energies around 1\,TeV the flux could be detectable by CTA and just marginally by other IACTs. However, the whole cloud is very extended and the detection of a more compact source ($\sim 0.1 \times 20$\,pc), such as the scenario considered above of reaccelerated/fresh accelerated cosmic rays is more feasible. The redistribution in energy and the hardening in the spectrum, suffered by the background cosmic ray protons  in the process of reacceleration and compression is fundamental for the detectability of the produced emission by IACTs.

\subsection{Adiabatic shock}\label{sec:as}

In the case of a fast shock, and/or in the case of low density, the condition given by Eq.\,(\ref{eq4}) is not fulfilled and the shock is adiabatic. This situation might occur in very tenuous clouds or in regions of low density inside the cloud as the actual density is expected to be inhomogeneous. Another possibility is when the shock velocity is very high, however shocks with $V > 300$\,km\,s$^{-1}$ would be rare in these sources. Also in the case of subsolar metallicities cooling times are expected to be longer and hence shocks might radiate less efficiently. The case of particles accelerated in  adiabatic shocks in HVCs induced by cloud-disk collisions has been extensively studied in \citet{2018MNRAS.475.4298D}. 

Aside for the differences in the modeling, the nonthermal emission that we obtained here from reaccelerated cosmic rays in HVCs is much higher than the one expected from freshly accelerated particles in an adiabatic shock \citep[see][]{2018MNRAS.475.4298D}. \citet{Inoue_2017} also model the nonthermal emission induced by HVCs colliding with the Galactic plane aiming to explain the origin of dark GeV-TeV sources; the radiation they obtained for a source at $d = 20$\,kpc and $V_{\rm s} = 300$\,km\,s\,$^{-1}$ is of similar order as the one obtained in this work from cosmic rays reaccelerated in a radiative shock of the same velocity.

\subsection{Wave damping}\label{sec:wd}

In the previous study we consider that the shock would provide sufficient heating for ionizing the entire preshock and postshock \citep{1996ApJS..102..161D}, hence preventing Alfvén wave damping by neutrals. The ionization degree (before the shock passage) of HVCs is not easy to measure. Measurement of both HI and H$\alpha$ emission together with the distance allows a derivation of
the volume density and of the ionization fraction. \citet{2008ApJ...672..298W} computed ionization fractions for clouds of known distance, finding ionized gas mass factors $1-3$ larger than the mass of neutral gas. In general, the standard model for HVCs consists of an outer part of completely ionized material and a neutral core \citep{2013pss5.book..587W}. If the interaction occurs in the external region, plus the ionizing effect from the shock, wave damping by neutrals can be negligible. The region of interaction can be account by a filling factor $f = \zeta^3$ (this is considered in the results above as well, because we do not assume that the whole cloud is interacting with the shock), then is not the whole source participating in the reacceleration or acceleration of the cosmic rays but  a fraction $f$. 

In order to quantify the degree of wave damping we can estimate the ion-neutral damping rate $\Gamma$ \citep[e.g.,][]{1996A&A...309.1002O}:

\begin{eqnarray}
\Gamma = \begin{cases}
8.4 \times 10^{-9}\,n_{\rm H}\left(\frac{T_{\rm c}}{10^4\,{\rm K}}\right)^{0.4}\left(\frac{\kappa}{\kappa_{\rm c}}\right)^2\,{\rm s}^{-1} & \quad \kappa  > \kappa_{\rm c}, \\
8.4 \times 10^{-9}\,n_{\rm H}\left(\frac{T_{\rm c}}{10^4\,{\rm K}}\right)^{0.4}\,{\rm s}^{-1} & \quad \kappa < \kappa_{\rm c},
\end{cases}
\end{eqnarray}
\noindent here $n_{\rm H}$ is the number density of neutrals and $\kappa_{\rm c} = \frac{\nu_{\rm io}}{V_{\rm A}}\frac{n_{\rm i}}{n_{\rm H}}$, with $\nu_{\rm io}$ the ion-neutral collision frequency and $V_{\rm A}$ the Alfvén velocity. The HVCs temperature is obtained from observations of optical emission lines and is of the order of $10^{4}$\,K \citep{2013pss5.book..587W}. For the values we are considering the wave-number $\kappa_{\rm c}$ corresponds to a particle resonant energy $E \sim 0.18$\,TeV. Above this energy the wave damping becomes inefficient. Following \citet{2016A&A...595A..58C}, in order to assess whether efficient acceleration (or reacceleration) is possible we compare the timescale for ion-neutral damping of turbulence at wavelength $\kappa_{\rm c}$ with the interaction time $t_{\rm int}$ for which the shock has been interacting with the cloud, i.e. $\Gamma\times t_{\rm int} < 1$, which yields an upper limit for the density of neutrals: $n_{\rm H} < 3.8\times 10^{-3}$\,cm$^{-3}$.   

The above upper limit for the density of neutrals is rather low, in this situation wave damping is important and shock acceleration might be inefficient. In this case the particles from the cosmic ray background experience only compression by the shock. The result is already presented in Figures \ref{v100} and \ref{v300} (blue line) for the two compression factors $S = 20$ and 60, corresponding to $V_{\rm s} = 100$ and 300\,km\,s$^{-1}$, respectively. For the slower shock the maximum gamma-ray luminosity, reached at $E \sim 0.4$\,GeV, is $10^{32}$\,erg\,s$^{-1}$ and it is 2.7 higher for the faster shock. This emission is undetectable for a IACT; at $E \sim 0.4$\,GeV $F_{\gamma} = 8.8\times 10^{-10}$\,cm$^{-2}$\,s$^{-1}$, and might be detectable by {\it Fermi}. 

Not only the acceleration efficiency might be affected by Alfvén wave damping but the confinement of the particles, that might lead to lower gamma-ray fluxes. Taking these effects into consideration is out of the scopes of the present study, but it is being considered in a forthcoming work. Another scenario that is worth discussing is the case in which background or fresh particles are accelerated in a HVC by magnetic reconnection. This situation is discussed in what follows.   

\subsection{Magnetic reconnection}
 
Magnetic reconnection can occur in HVCs when interacting with the magnetic field of the surrounding medium as proposed by \citet{1997A&A...320..746Z}. This process can heat the plasma to very high temperatures, producing X-rays \citep{2014ApJ...791...41H}. Magnetic reconnection has been proposed as an efficient mechanism for particle acceleration, and particles can undergo a first order Fermi process by scattering between the two opposed-polarized  converging flows \citep[e.g.,][]{2005A&A...441..845D,2012PhRvL.108x1102K,2016MNRAS.463.4331D,2021ApJ...908..193M}. In the presence of turbulence this mechanism becomes fast, hence being  more efficient for accelerating particles. As pointed out earlier, turbulence is expected in HVCs \citep[][]{2013pss5.book..587W}. 

The available power for particle acceleration comes from the kinetic energy of the system; the converging flows have velocity $V_{\rm rec}$, which in the case of turbulent reconnection is a fraction $\phi$ of the Alfvén velocity $V_{\rm A} = 100/(n_{\rm c}/{\rm cm}^{-3}) \times \left({B}/{50\,\mu{\rm G}}\right)$\,km\,s$^{-1}$ \citep{1999ApJ...517..700L}. Here we consider a higher value for the magnetic field in the cloud in order to maximize the effects of magnetic reconnection; we use $B_{\rm c} = 50\,\mu$G, which does not contradict estimations by observations (which are lower limits, see Sect.\,\ref{modelling}). For $\phi = 0.3$ and $R_{\rm c} = 20$\,pc the available power is $P_{\rm rec} = 0.5\,4{\pi} {\rho}  R_{\rm c}^{2}\,V_{\rm rec}^{3} \sim 1.5\times 10^{36}$\,erg\,s$^{-1}$.   

As a  first order Fermi process the resulting distribution is a power-law on the particles energy with index $\alpha_{\rm m}$. Some authors argued that $\alpha_{\rm m}$ is related to the compression factor, and that large values of $S$ are expected \citep[e.g.,][]{2012MNRAS.422.2474D}. However, as discussed in previous sections the presence of magnetic fields limits the compression factor. In fact, as the reconnection velocity is always sub-alfvenic the gas compression will always be low (the ram pressure is always smaller than the magnetic pressure), see for example Eq.(\ref{S}). The theoretical and/or numerical estimations of the power-law index  range from $\sim 5/2$ \citep[][]{2005A&A...441..845D}, $\sim 2$ \citep{2018MNRAS.481.5687P} to $\sim 1$ \citep{2012MNRAS.422.2474D,2016MNRAS.463.4331D}. { We} adopt the value   $\alpha_{\rm m} = 2$. 

In order to have an estimation of the expected radiation we compute the emission produced by the distribution of accelerated particles, mimicking what was done for shock acceleration in Sect.\,\ref{acc}. Freshly accelerated particles will then follow a power-law\footnote{Cosmic rays from the background Sea might as well undergo reacceleration in the magnetic reconnection process, we are not considering this case here.}
\begin{equation}
f_{\rm m} = {K}_{\rm m} \left(\frac{p}{p_{\rm inj}}\right)^{-\alpha_{\rm m}-2},
\end{equation}  
\noindent with  $p_{\rm inj}$ the injection momentum.  ${K}_{\rm m}$ is the normalization factor such that
\begin{equation}
K_{\rm m} = \frac{\eta_{\rm m} \rho_{\rm c} V_{\rm rec}^{2}}{4/3 \pi c \int \left(p/p_{\rm inj} \right)^{-\alpha_{\rm m}-2}p^3 \beta_p {\rm d}p},
\end{equation} 
\noindent here $\eta_{\rm m}$ is the efficiency of particle acceleration in the magnetic reconnection process. We solve then Eq.\,(\ref{trans}) using $f_{\rm m}$ as done for the freshly shock-accelerated cosmic rays for electrons and protons with $a = 100$. We use Eq.\,(\ref{emax2}) for estimating the maximum achievable energy, but for an acceleration rate related to the reconnection velocity $V_{\rm rec}$. Here we have to redefine the meaning of the interaction time (there is no shock-front moving through the source): we assume that the acceleration process occurs during a time $t_{\rm int}$ which is a fraction $\zeta$ of a typical dynamical time scale of the system $t_{\rm dyn} \equiv R_{\rm c}/V_{\rm rec}$. This gives $E_{\rm max} = 27$\,TeV for protons. In the case of electrons the maximum energy is determined by synchrotron losses, and this gives $\sim$ 1.5\,TeV.

   \begin{figure}
   \centering
   \includegraphics[width=0.6\linewidth, trim=1.8cm 0cm 1.8cm 0cm, angle=270]{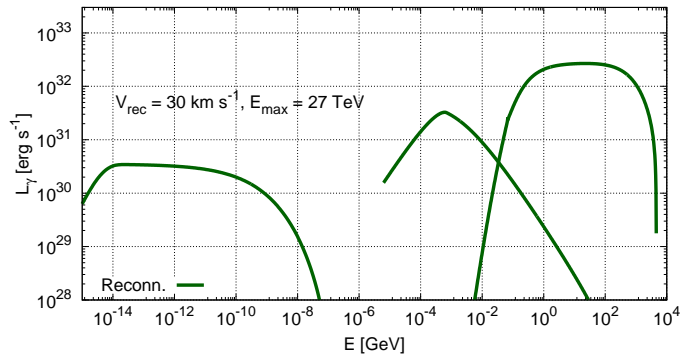}
      \caption{Nonthermal emission expected from the interaction of fresh particles accelerated by magnetic reconnection in a HVC.}
         \label{reconnection}
   \end{figure}
   
The resulted emission is plotted in Figure\,\ref{reconnection}. The emission is not as high as the case of a $100$\,km\,s$^{-1}$ shock, because even for a high magnetic field $V_{\rm rec} < V_{\rm s}$. However, the effects of magnetic reconnection might lead to the acceleration of particles in a shock, the emission produced by reconnection might help to photoinize the plasma, then diminishing the effects of wave damping by neutrals (see discussion above). Furthermore, during the process shocks are formed and are believed to be capable of accelerating particles \citep[e.g.,][]{1987SSRv...46..113S}. 

\subsection{Neutrinos}\label{neutrinos}

A by-product of proton-proton interactions are neutrinos (both muon and electronic neutrino). The amount of power that goes to neutrino can be estimated as a fraction of the produced gamma luminosity from neutral pion decay. This fraction is between 0.02 and 0.05 \citep[e.g.,][]{1992A&A...257..465A}. Here we make a rough estimation of the total neutrino  upper limit fluxes expected from the whole population of HVCs (the expected flux from a single cloud is too low), based on our gamma-ray results. This estimation is an upper limit, giving that the background cosmic ray population used is the local one which is expected to be higher than that of the halo.

We obtained a proton-proton luminosity of $\approx$ $5\times10^{35}$\,erg\,s$^{-1}$, for a $M = 10^{5}$\,M$_{\odot}$ cloud. The total neutral HVC gas mass is $\approx$ $2.5\times 10^{8}$\,M$_{\odot}$ \citep{2004ASSL..312.....V,2005A&A...432...45B}. Hence the total mass is expected to be a factor of 2 higher: $M_{\rm HVC} = 5\times10^8$\,M$_{\odot}$. Then an upper limit for the gamma-ray emission expected from HVCs as passive sources is $L_{\rm HVC} = 2.5\times10^{39}$\,erg\,s$^{-1}$. Consequently, an upper limit for the power in neutrinos produced through proton-proton interaction is $P_{\nu\,,{\rm HVC}} \approx 1.25\times 10^{38}$\,erg\,s$^{-1}$. 

Considering that more than 60\% of this mass comes from the Magellanic Stream located at $d = 55$\,kpc \citep{2017ASSL..430...15R}, the expected flux is $3.5\times10^{-10}$\,cm$^{-2}$\,s$^{-1}$ at 1 TeV and much lower at energies above $100$\,TeV where high energy neutrino detectors, such as IceCube, operate. The expected neutrino upper limit flux is extremely low, even when considering nearby clouds. 

\section{Summary and conclusions}\label{conclusion}

In this paper we analyze the scenario in which cosmic rays interact within shocked HVCs, aiming to estimate the produced nonthermal emission. We consider the case in which particles from the cosmic-ray Sea:  protons and electrons plus locally produced pairs, are reaccelerated in radiative shocks formed within the cloud as a result of the collision with the Galactic disk and reenergized by compression. We also study the case in which fresh particles are accelerated at the shock. We make very general assumptions, without considering extreme situations/parameters. An extensive discussion is presented on many possible scenarios depending on the clouds' physical conditions.\\

We arrived at the following conclusions:

\begin{itemize}

\item The presence of shocks enhances largely the population of cosmic rays in the cloud, by reacceleration and compression; the hardening of the proton distribution is key for detectability with IACTs. 

\item The maximum energy achievable by the particles in both the acceleration and reacceleration scenarios is a determination factor in the production of nonthermal emission. This energy depends directly on the diffusion regime of the high-energy particles.

\item The leptonic contribution to the SED is important at radio wavelengths and soft gamma rays.

\item A shock with $V_{\rm s} = 100$\,km\,s$^{-1}$ produces a compression factor $\sim$ 20 in our model, and the maximum acceleration energies are between $4$\,GeV and 9\,TeV. The highest maximum energy case produces higher luminosities, peaking at $\sim$ $6\times 10^{33}$\,erg\,s$^{-1}$ { at $\sim$ 100\,GeV}; the synchrotron luminosity peaks at radio wavelengths with $2\times 10^{33}$\,erg\,s$^{-1}$. 

\item For a faster shock of  $V_{\rm s} = 300$\,km\,s$^{-1}$ the compression factor is $\sim$ 60, the achievable maximum energies are between $36$\,GeV and 27\,TeV. The highest gamma luminosity is $\sim$ $	1.7\times 10^{34}$\,erg\,s$^{-1}$ for $E_{\gamma} \sim 100\,$GeV. This is the most promising detectable scenario among the radiative shock cases studied here; in this case the maximum produced radio synchrotron luminosity is $2\times 10^{34}$\,erg\,s$^{-1}$.

\item Secondary pairs dominate the SED for the higher maximum energy cases, while reaccelerated electrons from the background and freshly accelerated electrons dominate the leptonic SED in the lower maximum energy cases.

\item Significant nonthermal X-ray emission is produced in the case with  $V_{\rm s} = 300$\,km\,s$^{-1}$ and $E_{\rm max}=$ 27\,TeV. However, given the spatial extension of the source a detection with current X-ray satellites might be difficult.

\item In the absence of efficient shock acceleration (in the case of low ionization, for example) only compression would re-energize the cosmic rays. The maximum gamma-ray luminosity produced by the compressed cosmic ray protons is $\sim$ $2.7\times10^{32}$\,erg\,s\,$^{-1}$;

\item Magnetic reconnection might accelerate fresh particles and cosmic rays from the background. Furthermore, the process could improve the scenario for particle acceleration by ionizing the material through the emission produced in the reconnection event and by the creation of shocks where particles can also be accelerated; 

\item The obtained gamma-ray emission from a $10^5$\,$M_{\odot}$ passive cloud is $\sim$ $10^{35}$\,erg\,s$^{-1}$;

\item We estimate the upper limit of power in neutrinos expected from the population of HVCs via proton-proton interactions as passive sources, giving an undetectable value of $P_{\nu\,,{\rm HVC}} \approx 1.25\times 10^{38}$\,erg\,s$^{-1}$. 

\item It is more favorable, from the observational point of view, the scenario in which an enhanced population of background cosmic rays is producing the emission in a fraction of the cloud, than the whole cloud acting as a passive target for the cosmic-ray Sea.

\item Provided that the source is relatively nearby some of the cases considered { in this work} would produce luminosities above the sensitivities of IACTs, and by {\it Fermi} in some cases. 
\end{itemize}

\section*{Acknowledgements}
The author thanks F.~L. Vieyro and R. Santos-Lima for carefully reading the manuscript and for useful discussions. This work is supported by the Brazilian Agency FAPESP through the Grants 2019/05757-9 and 2020/08729-3. 

\section*{Data availability}
The data underlying this article will be shared on reasonable request to the corresponding author.



\bibliographystyle{mnras}
\bibliography{bibliography} 





\bsp	
\label{lastpage}
\end{document}